\documentclass[prb,aps,twocolumn,showpacs]{revtex4}%
\usepackage{amsmath}
\usepackage{amsfonts}
\usepackage{amssymb}
\usepackage{graphicx}%
\usepackage{bm}
\begin{document}
\title{Superfluid density and specific heat within 
self-consistent scheme for two-band superconductor}
\author{V. G. Kogan}
\email{kogan@ameslab.gov}
\affiliation{Ames Laboratory and Department of Physics \& Astronomy, Iowa State University, Ames, IA 50011}
\author{C. Martin}
\email{cmartin@ameslab.gov}
\affiliation{Ames Laboratory and Department of Physics \& Astronomy, Iowa State University, Ames, IA 50011}
\author{R. Prozorov}
\email[Corresponding author: ]{prozorov@ameslab.gov}
\affiliation{Ames Laboratory and Department of Physics \& Astronomy, Iowa State University, Ames, IA 50011}

\date{30 April 2009}

\begin{abstract}

The two gaps in a two-band clean s-wave superconductor are evaluated self-consistently within the quasiclassical Eilenberger weak-coupling formalism  with two in-band and one inter-band pairing potentials. Superfluid density, free energy and specific heat are given in the form amenable for fitting the experimental data. Well-known two-band MgB$_2$ and V$_3$Si superconductors are used to test the developed approach. The pairing potentials obtained from the fit of the superfluid density data in MgB$_2$ crystal were used to calculate temperature-dependent specific heat, $C(T)$. The calculated $C(T)$ compares very well with the experimental data. Advantages and validity of this, what we call the ``$\gamma$-model", are discussed and compared with the commonly used empirical (and \textit{not self-consistent}) ``$\alpha$-model". Correlation between the sign of the inter-band coupling and the  signs of the two order parameters is discussed. Suppression of the critical temperature by the inter-band scattering is evaluated and shown to be severe for the inter-band repulsion as compared to the attraction. The data on a strong $T_c$ suppression in MgB$_2$ crystals by impurities suggest that the order parameters on two effective bands of this material may have opposite signs, i.e., may have the $s_{\pm}$ structure similar to the current proposals in iron-based pnictide superconductors.

\end{abstract}

\maketitle

\section{Introduction}

Nearly all superconductors discovered recently and some
 well-studied compounds (e.g., Nb$_2$Se and
V$_3$Si) are multiband materials with complex Fermi surfaces
and anisotropic order parameters. Measuring temperature
dependences of the London penetration depth  $\lambda(T)$, often converted to the superfluid density  $\rho(T)= \lambda(0)^2/\lambda(T)^2$, and of the electronic specific heat  $C(T)$, are among primary tests directly linked to pairing mechanisms of new superconductors. Still, the
methods employed to interpret the data are often empiric with
simplicity as a main justification.

The most popular among practitioners ``$\alpha$-model" takes a shortcut by assigning the BCS temperature dependence to the two gaps $\Delta_{1,2}$ with which to fit  data on the specific heat \cite{Bouquet01} and the superfluid density $\rho=x\rho_1+(1-x)\rho_2$.\cite{fletcher2005,Prozorov2006} 
Here, $\rho_{1,2}$ are evaluated with 
$\Delta_{1,2}=(\alpha_{1,2}/1.76)\Delta_{BCS}(T)$ and $x$ takes account the relative band contributions. Although the $\alpha$-model had played an important and timely role in providing convincing evidence for the  two-gap
superconductivity in MgB$_2$, \cite{Bouquet01,fletcher2005}  
it is intrinsically inconsistent in the most important task of this procedure, namely, in describing properly the temperature dependencies of $\rho(T)$ and $C(T)$. In fact, one cannot \textit{a priory} assume temperature dependencies for the gaps in the presence of however weak interband coupling imposing the same $T_c$ for two bands. In unlikely situation of zero interband coupling, two gaps would have single-gap BCS-like $T-$ dependencies, but will have two different transition temperatures, see Fig.\,\ref{fig1}.

The full-blown microscopic approach based on the Eliashberg theory, 
on the other hand, is quite involved and difficult for the data
analysis.~\cite{brinkman2002,Golubov} Hence, the need for
a relatively simple but justifiable, self-consistent and effective
scheme experimentalists could employ. The weak-coupling model is
such a scheme. Over the years, the weak-coupling  theory had proven
to describe well multitude of superconducting phenomena. Similar  to the weak coupling is the ``renormalized BCS" model
of Ref.\,\onlinecite{Carbotte} that incorporates the Eliashberg
corrections in the effective coupling constants in a manner
described below. We will call our approach a ``weak-coupling
two-band scheme" and clarify in the text below that 
applicability of the model for the analysis of the superfluid density
and specific heat data is broader than the traditional weak 
coupling.

The s-wave weak-coupling multigap model has been proposed at the dawn 
of superconductivity theory by Moskalenko \cite{M59} and Suhl,   Matthias, and   Walker\cite{SMW}  when numerical tools
were still in infancy. In this work we basically follow these seminal
publications to develop a self-consistent procedure for the penetration
depth data analysis. Our scheme allows one to connect
between two independent data sets, the superfluid density and
the specific heat, thus providing a reliability check upon values of the coupling constants extracted from fitting the data. We also discuss the suppression of the critical temperature $T_c$ by non-magnetic impurities and suggest that the data on this suppression for MgB$_2$ are consistent with a weak repulsive interband interaction that corresponds to opposite signs of the order parameter on two bands, i.e., to $\pm\,$s structure of the order parameter. 

To test our formal scheme, the data for two known s-wave two-gap superconductors, MgB$_2$ and V$_3$Si, were used. The specific heat data were taken from the Ref.\,\onlinecite{Wang2001}. The magnetic penetration depth, $\lambda(T)$, was measured by using a self-oscillating tunnel-diode resonator (TDR) with resonant frequency $f_{0}  \approx 14$ MHz. The measured quantity is the shift of this frequency, $ f(T)-f_{0} =-4 \pi\chi(T){\cal G}$, where $\chi$ is the total magnetic susceptibility in the Meissner state and ${\cal G}\simeq f_{0}V_{s}/2V_{c}\left( 1-N\right) $ is a geometric  factor defined by volumes of the coil, $V_c$, and of the sample,  $V_{s}$,  and by the demagnetization factor $N$. ${\cal G}$ is measured directly by pulling the sample out of the coil at the lowest temperature. \cite{Prozorov2006}  For the susceptibility  we use   $4\pi\chi\left( T\right) =\lambda(T)/w\,\,\tanh[w/\lambda(T)]-1$. \cite{Prozorov2000} 

\section{Eilenberger two-band scheme}

  Perhaps, the simplest formally weak-coupling approach is based on
the Eilenberger quasiclassical formulation of the superconductivity
 valid for general anisotropic order parameters and  Fermi 
surfaces.\cite{E} Eilenberger functions $f,g$ for clean materials 
in equilibrium obey the system:
\begin{eqnarray}
0&=&2\Delta g/\hbar  -2\omega f \,,\label{eil1}\\
g^2&=&1-f^2\,, \label{eil3}\\
\Delta( {\bm k})&=&2\pi TN(0) \sum_{\omega >0}^{\omega_D} \Big\langle
V({\bm k},{\bm k}^{\prime\,}) f({\bm k}^{\prime},\omega)\Big\rangle_{{\bm k}^{\prime\,}}.
   \label{self-cons}
\end{eqnarray}
Here, ${\bm k}$ is the Fermi momentum; $\Delta$ is the gap
function which may depend on the position ${\bm  k}$ at
the Fermi surface   in cases other than the isotropic
s-wave.    Further,
$N(0)$ is the total density of states at the Fermi level per one spin;
the   Matsubara frequencies  are defined by $\hbar\omega=\pi
T(2n+1)$ with an integer $n$, and $\omega_D$ is the Debye
frequency; $\left\langle...\right\rangle$ stands for averages over
the Fermi surface.
% weighted with the local density of states $
%\propto 1/|{\bm v}|$ are defined as
%\begin{equation}
%\Big\langle X \Big\rangle = \int \frac{d^2{\bm  k}}
%{(2\pi)^3\hbar
%N(0)|{\bm v}|}\,\,X\,.\label{<>}\end{equation}
As  a {\it weak coupling} theory the Eilenberger scheme deals with the
{\it effective electron-electron coupling} $V$ responsible for
superconductivity;  properties of
intermediate bosons (phonons  or other possible mediators)   
enter   via properly renormalized  $V$.

   Consider a model material with the gap given by
\begin{equation}
\Delta ({\bf k})= \Delta_{1,2}\,,\quad {\bf k}\in   F_{1,2} \,, 
 \label{e50} 
\end{equation}
where $F_1,F_2$ are two sheets of the Fermi surface. 
 The gaps are constant at each band. Denoting
the densities of states on the two parts as $N_{1,2}$, we have 
 for a quantity $X$ constant at each Fermi sheet:
 \begin{equation}
\langle X \rangle = (X_1 N_1+X_2 N_2)/N(0) = n_1X_1+
n_2X_2\,,
\label{norm2}
\end{equation}
where $n_{1,2}= N_{1,2}/N(0)$; clearly, $n_1+n_2=1$. 

  Equations (\ref{eil1}) and (\ref{eil3})
 are easily solved.  Within the two-band model we have:
\begin{equation}
f_\nu  ={\Delta_\nu  \over
\beta_\nu },\quad g_\nu   =  {\hbar\omega\over
\beta_\nu } ,\quad
\beta_\nu ^2=\Delta_\nu ^2+ \hbar^2\omega^2\,, \label{f_n}
\end{equation}
where the band index $\nu=1,2$. 
The  self-consistency equation (\ref{self-cons}) takes the form:
\begin{eqnarray}
\Delta_\nu= 2\pi T\sum_{ \mu=1,2} n_\mu \lambda_{\nu\mu} 
f_\mu 
=  \sum_{\mu} n_\mu \lambda_{\nu\mu} 
\Delta_\mu \sum_{\omega }^{\omega_D}\frac{2\pi T}{\beta_\mu} ,
   \label{self-cons1}
\end{eqnarray}
where $\lambda_{\nu\mu} = N(0)V(\nu,\mu)$ are dimensionless 
effective interaction constants. 

A remark is here in order on applicability of Eq.\,(\ref{self-cons1})
 central for our approach. Starting with the general
Eliashberg formalism, Nicol and Carbotte derived a ``renormalized
BCS" self-consistency equation:\cite{Carbotte}
\begin{eqnarray}
\Delta_\nu&=& \frac{2\pi T}{Z_\nu}\sum_{ \mu,\,\omega}^{\omega_D}  
\left(\lambda_{\nu\mu}^{ph}-\mu^*_{\nu\mu}\right)  f_\mu \, ,
   \label{nicol}
\end{eqnarray}
where $\lambda_{\nu\mu}^{ep}$ is the  coupling due to
electron-phonon interaction, 
$\mu^*_{\nu\mu}$ describes the Coulomb interaction, and
$Z_\nu=1+\sum_\mu \lambda_{\nu\mu}^{ep}$ is the strong coupling
renormalization. Replacing
$\left(\lambda_{\nu\mu}^{ep}-\mu^*_{\nu\mu}\right)/Z_\nu \to n_\nu
\lambda_{\nu\mu}$
we obtain our Eq.\,(\ref{self-cons1}). One should have this in mind
while interpreting the constants $\lambda_{\nu\mu}$ which can be
obtained from fitting the data to our ``renormalized 
weak-coupling" model.  

 Note that the notation commonly
used   in literature for $\lambda^{(lit)}_{\nu\mu}$ differs from
ours: $\lambda^{(lit)}_{\nu\mu}=n_\mu \lambda_{\nu\mu}$. We find
our notation convenient since, being proportional to the coupling
potential, our coupling matrix is symmetric:
$\lambda_{\nu\mu}=\lambda_{\mu\nu}$.  
It is worth stressing that for a given coupling
matrix $\lambda_{\mu\nu}$, relative densities of states 
$n_\nu$, and the energy scale $\hbar\omega_D$, 
Eq.\,(\ref{self-cons1})  determines both
$T_c$ and $\Delta_{1,2}(T)$. 
%We are interested here only in $T_c$ and$\Delta_{1,2}(0)$. 

 \subsection{Critical temperature $\bm T_c$}

\begin{figure}[tb]
\includegraphics[width=8.5cm]{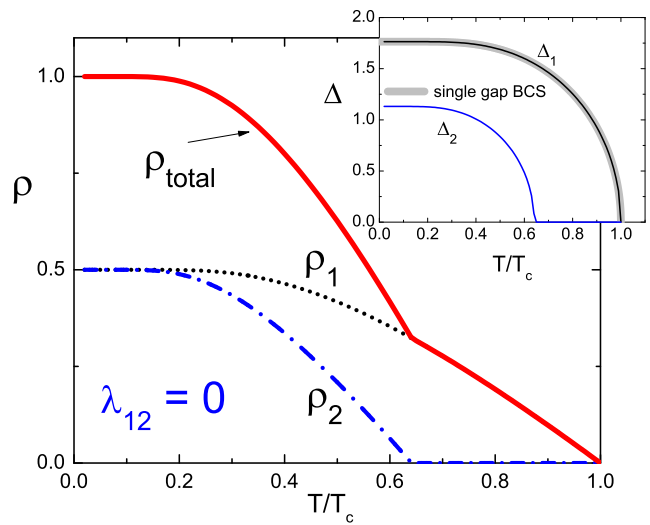}
\caption{(Color online) Calculated superfluid density and  the gaps vs. reduced 
temperature (inset) for zero interband coupling, $\lambda_{12}=0$. In this calculation  $\lambda_{11}=0.5$,  $\lambda_{22}=0.45$, $n1=n2=0.5$, and $\gamma=0.5$, see text below.}
\label{fig1}
\end{figure}

As $T\to T_c$,   $\Delta_{1,2}\to 0$, and $\beta\to\hbar\omega$. 
The sum over $\omega$ in Eq.\,(\ref{self-cons1}) is readily evaluated:
\begin{eqnarray}
S=  \sum_{\omega }^{\omega_D}\frac{2\pi T}{\hbar\omega} = \ln\frac{2\hbar\omega_D}{T_c\pi e^{-\gamma}}=  \ln\frac{2\hbar\omega_D}{1.76\,T_c }, 
   \label{S}
\end{eqnarray}
 $\gamma=0.577$ is the Euler constant.  This relation can also be written as
 \begin{eqnarray}
1.76\,T_c=  2\hbar\omega_D e^{-S}\,. 
   \label{S-Tc}
\end{eqnarray}
 The system (\ref{self-cons1}) is linear and homogeneous:
\begin{eqnarray}
\Delta_1&=&  S (n_1 \lambda_{11} \Delta_1+ n_2 \lambda_{12} \Delta_2)\,,\nonumber\\
\Delta_2&=&  S (n_1 \lambda_{12} \Delta_1+ n_2 \lambda_{22} \Delta_2) .
   \label{systemTc}
\end{eqnarray}
It has nontrivial solutions $\Delta_{1,2}$  if its determinant
is zero:
\begin{eqnarray}
 S^2 n_1n_2\eta &-& S (n_1 \lambda_{11}   + n_2 \lambda_{22})+1=0 ,\nonumber\\
 \eta &=&  \lambda_{11}  \lambda_{22}-\lambda_{12}^2\,.
    \label{det=0}
\end{eqnarray}
The roots of this equation  are:
\begin{eqnarray}
&&  S=\frac{ n_1 \lambda_{11}   + n_2 \lambda_{22} \pm\sqrt{ (n_1 \lambda_{11} +n_2 \lambda_{22})^2 -4n_1n_2\eta}}{2n_1n_2\eta}, \nonumber\\ 
&&= \frac{ n_1 \lambda_{11}   + n_2 \lambda_{22} \pm\sqrt{ (n_1 \lambda_{11} - n_2 \lambda_{22})^2 +4n_1n_2\lambda^2_{12}}}{2n_1n_2\eta}. \nonumber\\ 
 \label{S1}
\end{eqnarray}
Since $ T_c\ll \hbar\omega_D$, Eq.\,(\ref{S-Tc}) shows that only positive $S$  are admissible. 
The second form (\ref{S1}) shows that both roots are real; their product is $S_1S_2=1/n_1n_2\eta$.
Choosing a proper root, one should consider various  
possibilities.  If $\eta >0$ (that implies  both $\lambda_{11}$ and 
$\lambda_{22}$ are of the same sign), Eq.\,(\ref{S1}) shows that
both roots are positive, and  one should choose the smallest (to
have maximum $T_c$). If $\eta <0$  ($ \lambda_{12}^2 > \lambda_{11}  \lambda_{22}$ that can happen (i) for a sufficiently strong interband coupling 
for both $\lambda_{11} $ and $\lambda_{22} $ positive or (ii) if
one of  $\lambda_{11} $,$\lambda_{22} $ is repulsive),  one
should   take the square root with  {\it minus}. It is of interest  to note that even for $\lambda_{11}=\lambda_{22} =0$, the interband coupling of either
sign may lead to superconductivity. In fact, $S = 1/\sqrt{
n_1n_2}\,|\lambda_{12}|$ for a
dominant  interband interaction $
|\lambda_{12}|\gg |n_1\lambda_{11} + n_2\lambda_{22}| $. 
However exotic, this possibility should not be ignored. 
This situation has been considered time ago by Geilikman
\cite{BTG} who found that interband Coulomb repulsion could lead to
superconductivity; recently this possibility has been considered by
Mazin and Schmalian in a discussion of superconductivity in the 
iron-pnictides.\cite{Maz} If $\eta=0$,   Eq.\,(\ref{det=0}) yields $S=1/(n_1
\lambda_{11}   + n_2 \lambda_{22})$. Finally, if the interband coupling is exactly zero, a quite unlikely situation, the second form of $S$ in Eq.\,(\ref{S1})
gives two roots $ 1/n_1\lambda_{11}$ and  $ 1/n_2\lambda_{22}$.
The smallest one gives $T_c$, whereas the other corresponds to
the temperature at which the small gap turns zero. This situation
is depicted in Fig.\,\ref{fig1}. 

We conclude this incomplete list  of possibilities by noting 
that within this model, interband  coupling enters $S$ only as
$\lambda^2_{12}$, i.e., interband attraction in clean materials affects $T_c$ exactly as does the repulsion. This is no longer true in the presence of interband scattering, the question discussed below. Denoting the properly chosen root as $S=1/{\tilde\lambda}$ we have:
\begin{eqnarray}
1.76\,T_c=  2\hbar\omega_D \exp (-1/{\tilde\lambda})\,.
   \label{Tc}
\end{eqnarray}
One easily checks that for all $\lambda$'s equal this yields the 
standard BCS result. 
 Among various possibilities we mention here   the case  
$\eta=\lambda_{11}\lambda_{22}-\lambda_{12}^2=0$ for which  
\begin{eqnarray}
 {\tilde\lambda}=n_1\lambda_{11}+n_2\lambda_{22}=
\langle\lambda\rangle \,.
   \label{Tc4}
\end{eqnarray}
This case corresponds to a popular model with factorizable
coupling potential $V({\bm k},\bm k^\prime)=V_0\Omega(\bm
k)\Omega(\bm k^\prime)$.\cite{Kad} This potential is amenable for
analytic work, but it  reduces severely the richness of the
two-band scheme.

Since the determinant of the 
system (\ref{systemTc}) is zero, the two equations are 
equivalent and  give near $T_c$: 
\begin{eqnarray}
\frac{\Delta_2}{\Delta_1}= \frac{ {\tilde\lambda}-n_1\lambda_{11}}{n_2\lambda_{12}} 
  \,.   \label{sign}
\end{eqnarray}
When the right-hand side is negative, $\Delta$'s are of opposite 
signs.  Within the   
one-band BCS, the sign of $\Delta $ is a matter of convenience; in
fact for one band the self-consistency equation determines only $|\Delta|$. For 
two bands, $\Delta_1$ and $\Delta_2$ may have opposite signs. If 
 the $\Delta$ values
  are $+D_1$ and $- D_2$, Eq.\,(\ref{self-cons1})  shows that  $-D_1$
and $+D_2$ is a solution too. One should be aware of this
multiplicity of solutions when solving the  system 
 (\ref{self-cons1}) numerically.  The problem is
even worse  because $\Delta_1=\Delta_2=0$ is always a solution. 

\begin{figure}[tb]
\includegraphics[width=8.5cm]{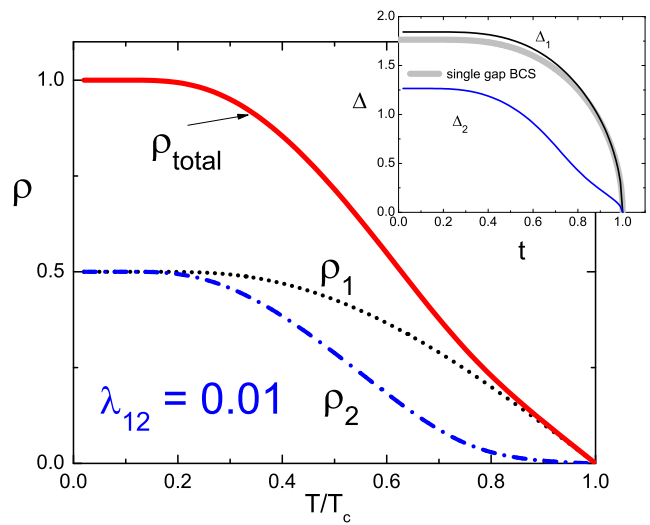}
\caption{(Color online) Calculated superfluid density and  the gaps vs. reduced 
temperature (inset) for  $\lambda_{12}=0.01$,  $\lambda_{11}=0.5$,  $\lambda_{22}=0.45$, $n1=n2=0.5$, and $\gamma=0.5$, see text below.}
\label{fig2}
\end{figure}

 \subsection{Order parameter }

Turning to evaluation of $\Delta_\nu(T)$, we note that the 
sum in Eq.\,(\ref{self-cons1}) is logarithmically divergent.  
 To deal with this difficulty, we employ  Eilenberger's idea of  
replacing $\hbar\omega_D$ with the measurable
$T_c$.  These are related by
Eq.\,(\ref{Tc}) which can be written as
\begin{eqnarray}
\frac{1}{{\tilde\lambda}}=  \ln\frac{T}{T_c}+\sum_{\omega }^{\omega_D}\frac{2\pi T}{\hbar\omega} ,
   \label{Eil0}
\end{eqnarray}
Now add and subtract the last sum from one in
Eq.\,(\ref{self-cons1}):
\begin{eqnarray}
&&\Delta_\nu=   \sum_{ \mu} n_\mu \lambda_{\nu\mu} \Delta_\mu \left[\sum_{\omega }^{\omega_D}\left(\frac{2\pi T}{\beta_\mu}- \frac{2\pi T}{\hbar\omega}\right) +  \sum_{\omega }^{\omega_D}\frac{2\pi T}{\hbar\omega}\right]\nonumber\\
&&=   \sum_{ \mu} n_\mu \lambda_{\nu\mu} \Delta_\mu \left[\sum_{\omega }^{\infty}\left(\frac{2\pi T}{\beta_\mu}- \frac{2\pi T}{\hbar\omega}\right) +\frac{1}{{\tilde\lambda}}- \ln\frac{T}{T_c}\right] \nonumber\\
   \label{self-cons2}
\end{eqnarray}
The last sum over $\omega$ is fast-converging and one can 
replace $\omega_D$ with $\infty$. Numerically, the upper limit of
summation over $n$ can
be set as a few hundreds that suffices even for low temperatures. Introducing dimensionless quantities
\begin{eqnarray}
\delta_\nu=  \frac{\Delta_\nu}{2\pi T} = 
 \frac{\Delta_\nu}{ T_c}\, \frac{1}{2\pi t}\,,
   \label{deltas}
\end{eqnarray}
with $t=T/T_c$, we rewrite Eq.\,(\ref{self-cons2}):
\begin{eqnarray}
 \delta_\nu= \sum_{ \mu=1,2} n_\mu \lambda_{\nu\mu} \delta_\mu
\left(  \frac{1}{{\tilde\lambda}}+\ln\frac{T_c}{T}
-A_\mu\right)\,, \nonumber\\
A_\mu =  \sum_{n=0}^{\infty}\left(\frac{1}{ n+1/2  }- \frac{1}
{\sqrt{\delta_\mu^2+(n+1/2)^2}} \right)\,.
   \label{.eps}
\end{eqnarray}
For given coupling constants $\lambda_{\nu\mu}$ and densities
of states $n_\nu$, this system of two equations can be solved
numerically for $\delta_\nu$ and therefore provide the gaps
$\Delta_\nu=2\pi T\delta_\nu(t)$. Two examples of these solutions with the sets of parameters differing only in $\lambda_{12}$ are given in insets to Figs.\,\ref{fig1} and \ref{fig2}. We observe that even a small interband coupling changes drastically the behavior of the small gap.

\subsection{Superfluid density}

Having formulated the way to evaluate $\Delta(T)$, we turn to the
London  penetration depth  given for general anisotropies of
the Fermi surface and of $\Delta$ by (see e.g.,
Ref.\,\onlinecite{Kogan2002}):
\begin{equation}
\left(\lambda_L^2\right)_{ik}^{-1}= \frac{16\pi^2 e^2N(0)T}{
c^2}\,  \sum_{\omega} \Big\langle\frac{
\Delta_0^2v_iv_k}{\beta ^{3}}\Big\rangle \,.  \label{lambda-tensor}
\end{equation}
where $v_i$ is the Fermi velocity. We consider here only the case
of currents in the $ab$ plain of uniaxial or cubic materials with
two separate Fermi surface sheets, for which a simple algebra gives
for the superfluid density
$\rho=\lambda_{ab}^2(0)/\lambda_{ab}^2(T)$:
\begin{eqnarray}
\rho  &=& \gamma {\rho _1} + \left( {1 - \gamma } \right){\rho
_2}\,,
\nonumber\\
\rho _\nu &=&  \delta_\nu^2\sum_{n=0}^\infty   
  \left[\delta_\nu^2+(n+1/2)^2\right]^{-3/2}\,,
\nonumber\\
\gamma  &=& \frac{{{n_1}v_{1}^2}}{{{n_1}v_{1}^2 + 
{n_2}v_{2}^2}}\,.
\label{rhogamma}
\end{eqnarray}
where $v_{\nu}^2$ are averages over corresponding band of the in-plane Fermi velocities. The formal similarity of the first line here to the widely used
$\alpha$-model (with an $x$ in place of out $\gamma$) prompts to name our scheme  
 the ``$\gamma$-model" (we do not renormalize the gaps, so no analog of $\alpha$ exists in our scheme. We note,
however, that our $\gamma$ that determines partial contributions from
each band is not just a partial density of states $n_1$ of the 
$\alpha$-model, instead it involves the band's Fermi velocities.

\begin{figure}[tb]
\includegraphics[width=8.5cm]{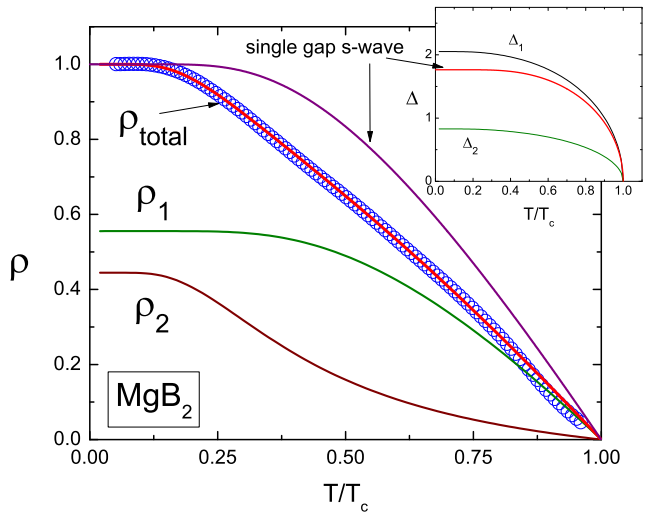}
\caption{(Color online) The data and fits of the superfluid density for MgB$_2$ single crystal and corresponding temperature-dependent gaps (inset). The fitting parameters are: $\lambda_{11}=0.23$, $\lambda_{22}=0.08$, $\lambda_{12}=0.06$;   the partial density of states   $n_{1}=0.44$ and $\gamma = 0.56$ were fixed.}
\label{fig3}
\end{figure}

We now apply the approach developed to fit the data for the superfluid density of MgB$_2$ crystals  acquired   by using the  TDR technique described above. Figure
\ref{fig3} shows the result of the fitting with three free
parameters:   $\lambda_{11}$, $\lambda_{22}$, 
and $\lambda_{12}$. The partial density of states and the parameter $\gamma$ were
taken from the the literature: the two-band mapping of the four-band MgB$_2$ gives  $n_1=0.44$   and the Fermi velocities $\langle v_{ab}^2\rangle_1=3.3$ and $\langle v_{ab}^2\rangle_2=2.3 \times 10^{15}\,$cm$^2$/s$^2$.\cite{Choi,Bel} 
This fit requires solving two coupled nonlinear equations, Eqs.~(\ref{.eps}). We used 
Matlab with the Optimization toolbox and utilized a nonlinear
solver using direct Nelder-Mead simplex search method. \cite{NM} 
The result is shown in Fig.\,\ref{fig3} with the
best fit parameters listed in the caption. We show below that the
same set of parameters used to calculate the free energy and the 
specific heat reproduces the data on  $C(T)$ remarkably well. 

\begin{figure}[tb]
\includegraphics[width=8.5cm]{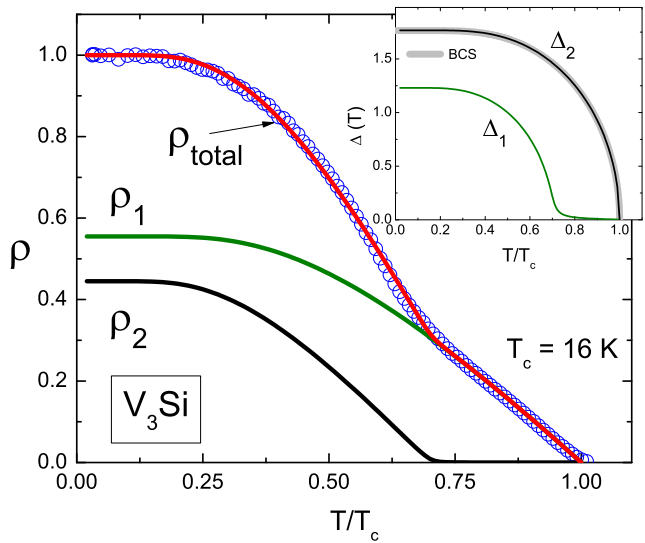}
\caption{(Color online) The data and fits of the superfluid density  for V$_3$Si single crystal and corresponding temperature-dependent gaps (inset). The fitting parameters are: $\lambda_{11}=0.1$, $\lambda_{22}=0.1$, $\lambda_{12}=1\times 10^{-5}$, $n_{1}=0.47$ and $\gamma = 0.4$.}
\label{fig4}
\end{figure}

Our numerical experimentation shows that if, in addition to coupling constants, the partial densities of states and the Fermi velocities are unknown, the fitting procedure becomes unstable  and good fits can be found for various combinations of the coupling constants. In the case of MgB${_2}$ these quanities are known and our fitting is quite certain.

For V$_3$Si we do not have detailed information regarding the band structure, partial
densities of states, and Fermi velocities on separate sheets of the Fermi surface of this material.  Hence, we took all those as free  parameters in the fitting procedure. The conclusions thus are less reliable for this material than for MgB$_2$: being mapped onto a two-band model, V$_3$Si comes out to have two nearly   decoupled bands with an extremely weak  interband coupling (still  sufficient  to give a single  $T_c$).  The results   and the best-fit parameters are given in the caption to Fig.\,\ref{fig4}. Note that the long linear tail in $\rho(t)$ as $T$ approaches $T_c$ is a direct manifestation of a very small gap, in this case $\Delta_1$, in this temperature domain.

\section{Free energy and specific heat}

By fitting the data for $\rho(t)$, we can extract the coupling
constants $\lambda_{\nu\mu}$ along with  $\Delta_\nu(T)$. This
allows one to determine all thermodynamic properties of the material
in question, of which we consider here  the specific
heat $C(T)$and its the jump  at $T_c$. To   this end, one  starts with the Eilenberger
expression for the energy difference:
\begin{eqnarray}
 \frac{F_n-F_s}{N(0)}&=&2\pi T   \sum_{ \nu,\omega} n_\nu  \frac{(\beta_\nu -\hbar\omega)^2}{\beta_\nu} .
   \label{energy}
\end{eqnarray}
Near $T_c$, one obtains:
\begin{eqnarray}
 \frac{F_n-F_s}{N(0)}&=&\frac{7\zeta(3)}{16\pi^2T_c^2} \sum_{ \nu } n_\nu  \Delta_\nu^4 
\nonumber\\
&=& 7\zeta(3) \pi^2T_c^2  \sum_{ \nu } n_\nu  \delta_\nu^4\, .
   \label{energy1}
\end{eqnarray}
Following Ref.\,\onlinecite{M59}, one can
look for solutions $\delta_\nu(t)$  of Eq.\,(\ref{.eps}) near
$T_c$ as an expansion:
 \begin{eqnarray}
 \delta_\nu = a_\nu \tau^{1/2}+b_\nu\tau^{3/2} \,,\qquad \tau=1-t .
   \label{expansion}
\end{eqnarray}
We substitute this in Eq.\,(\ref{.eps})  and compare terms
of different powers of $\tau$; the quantity $A_\mu=7\zeta(3)a_\mu^2\tau/2+{\cal
O}(\tau^2)$. In the lowest order we obtain the system of linear
homogeneous equations for
$a_{1,2}$ that coincides with the system (\ref{systemTc}). The same
arguments that led to Eq.\,(\ref{sign}) provide
\begin{eqnarray}
 a_2 = a_1 G = a_1 \frac{ {\tilde\lambda}-n_1\lambda_{11}} {n_2\lambda_{12}} \,.
   \label{sign1}
\end{eqnarray}
The conditions for existence of non-trivial solutions for 
$b_\nu$ in the next oder provide the second relation  for
 $a_\nu$. We omit a cumbersome algebra and give the result:
\begin{eqnarray}
 a_1^2 = \frac{2}{7\zeta(3)}\, \frac{{\tilde\lambda}^2-n_1n_2\eta}{
{\tilde\lambda}(n_1\lambda_{11}+n_2
\lambda_{12}G^3)-n_1n_2\eta} \, . 
   \label{a1}
   \end{eqnarray} 
\begin{figure}[tb]
\includegraphics[width=8.5cm]{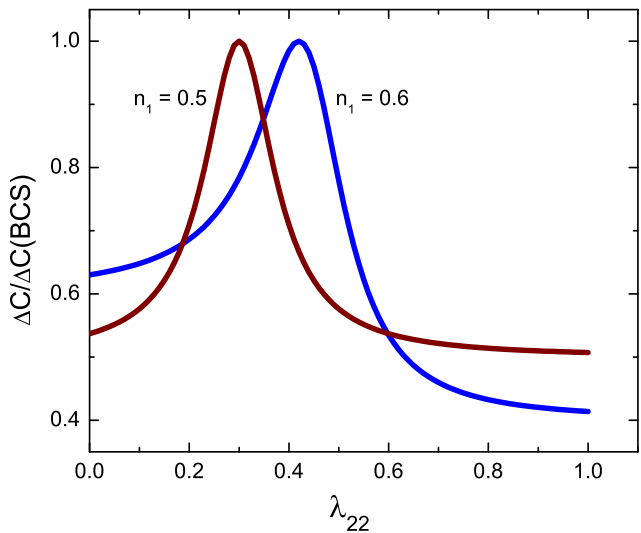}
\caption{(Color online) Dependence of the  specific heat jump on the mismatch between $\lambda_{11} $ and $\lambda_{22}$.  $\Delta C /C_N $ at $T_c$ normalized on the  BCS value of 1.43 is calculated using Eq.\,(\ref{jump1}) with $\lambda_{11}=0.3$, $\lambda_{12}=0.06$ and two values of $n_1$ and plotted {\it vs} $\lambda_{22} $. }
\label{fig5}
\end{figure}
We now obtain the energy near $T_c$, 
 \begin{eqnarray}
 F_n-F_s  = 7\zeta(3) N(0)\pi^2T_c^2  \sum_{ \nu } n_\nu  a_\nu^4\tau^2=B\tau^2\,,  
   \label{energy3}
\end{eqnarray}
and the specific heat:
 \begin{eqnarray}
 C_s-C_n=\frac{2B}{T_c}= 14\zeta(3)\pi^2 N(0)T_c   \sum_{ \nu } n_\nu  a_\nu^4\,.  
   \label{jump}
\end{eqnarray}
 The relative jump at $T_c$ is:
 \begin{eqnarray}
\frac{ \Delta C }{C_n}&=&\frac{12 }{7\zeta(3)}\,(n_1+n_2G^4)\nonumber\\
&\times&  \left[\frac{{\tilde\lambda}^2 -n_1n_2\eta}{
{\tilde\lambda}(n_1\lambda_{11}+n_2\lambda_{12}G^3)
-n_1n_2\eta}\right]^2.  
   \label{jump1}
\end{eqnarray}
If all coupling constants are the same, $\eta=0$, $G=1$, and
$\Delta C/C_n = 12 /7\zeta(3)=1.43$, as is should be. 
We note that the sign of the interband coupling $\lambda_{12}$ has no effect on the jump $ \Delta C$ since in Eq.\,(\ref{jump1}) $\lambda_{12}G$ is insensitive to this sign. 

It is easy to study numerically the jump dependence on the the three coupling parameters. As an example we show in Fig.\,\ref{fig5} the jump dependence on the mismatch between $\lambda_{11}$ and $\lambda_{22}$ for a fixed $\lambda_{12}$ for two values of $n_1$. One can see that for equal relative densities of states, the jump peaks at  $\lambda_{22}= \lambda_{11} $ at the one-band BCS value of 1.43; with the fixed  $\lambda_{11} $ and changing  $\lambda_{22} $ the jump drops with the mismatch  $|\lambda_{22}-\lambda_{11}|$. The peak position and the drop speed vary with varying bands contributions, so that the value of the jump {\it per se} cannot be interpreted as evidence for a particular order parameter.

\begin{figure}[t]
\includegraphics[width=8.5cm]{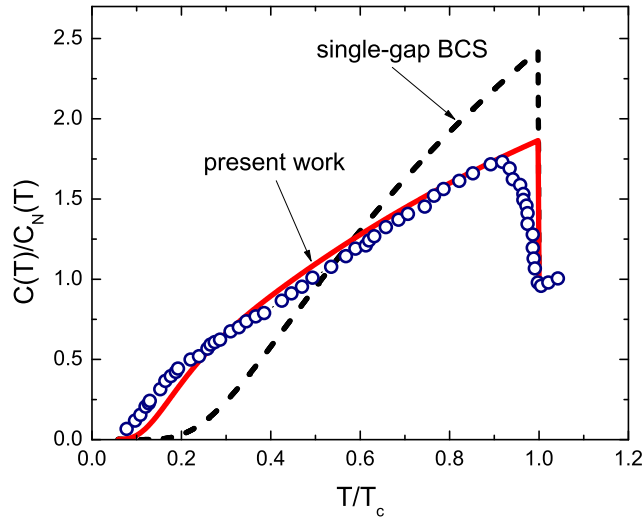}
\caption{(Color online) Normalized electronic specific heat, $C(T)/C_N(T)$. Symbols are the data from Ref.~\onlinecite{Wang2001}. The solid line shows result of the $\gamma$-model, it is calculated using the energy (\ref{energy})  with parameters determined from the fit to the data on superfluid density.}
\label{fig6}
\end{figure}

Now we can test our theory by employing the parameters 
obtained from the fit to the data on superfluid density as shown in
Fig.~\ref{fig3} and calculate the free energy and the
specific heat. The result is shown in Fig.\ref{fig6}. The dashed line
shows a single-gap weak-coupling BCS results and the  
solid line is the result of our calculations. Note that this is
not a fit, but a calculation with parameters  determined in independent measurement. The data shown by circles are taken from Ref.\,\onlinecite{Wang2001}. Since this is \textit{NOT} a fit, the agreement with the general behavior of $C(T)$ and, in particular, with value of the jump at $T_c$  is remarkable.

\section{$\bm T_c$ suppression by non-magnetic impurities}
The intraband scattering does not affect $T_c$, so that we focus on the
effect of interband scattering with an average scattering time $\tau$.
Since $g=1$ at $T_c$, the Eilenberger equations for $f_{1,2}$ in two
bands read:\cite{Kogan2004}
\begin{eqnarray}
0&=&2\Delta_1   -2\omega f_1+\nu_2(f_2-f_1)/\tau \,,\label{11}\\
0&=&2\Delta_1   -2\omega f_1+\nu_2(f_2-f_1)/\tau \,,\label{22} 
   \label{eil12}
\end{eqnarray} 
($\hbar=1$). This system yields:
\begin{eqnarray}
f_1&=&\frac{\Delta_1(\omega+n_1/2\tau) +
\Delta_2n_2/2\tau)}{\omega\omega^\prime}  
\,,\label{f1}\\ 
f_2&=&\frac{\Delta_2(\omega+n_2/2\tau) +
\Delta_1n_1/2\tau)}{\omega\omega^\prime}  
\,,\label{f2}
\end{eqnarray} 
where  $\omega^\prime=\omega+1/2\tau$. The self-consistency equation,  
\begin{eqnarray}
\Delta_\nu=   \sum_{ \mu,\omega } n_\mu \lambda_{\nu\mu} f_\mu\,,
\label{self-cons12}
\end{eqnarray}
again reduces to a system of linear and homogeneous equations for
$\Delta_{1,2}$, the determinant of which must be zero. Omitting the
algebra, we give the result:
\begin{eqnarray}
&& P^2 n_1n_2\eta - P (n_1 \lambda_{11}   + n_2
\lambda_{22}-n_1n_2\eta Q)\nonumber\\
&&+1-Q(n_1\eta_1+n_2\eta_2)=0
,\label{eqP}\\
&& \eta_1 =  n_1\lambda_{11} +n_2 \lambda_{12}\,,\quad \eta_2 = 
n_2\lambda_{22} + n_1 \lambda_{12}\, ,
    \label{etas}
\end{eqnarray}
 $\eta$ is defined in Eq.\,(\ref{det=0}). The quantities $P,Q$ are given
by:
\begin{eqnarray}
&& P = \sum_{\omega}\frac{2\pi T_c}{\omega^\prime} =
\ln\frac{\omega_D}{2\pi
T_c}-\psi\left(\frac{1}{2}+\frac{\rho_0}{2t}\right)\label{P}\\
&&Q = \frac{1}{2\tau}\sum_{\omega}\frac{2\pi T}{\omega\omega^\prime} =
\psi\left(\frac{1}{2}+\frac{\rho_0}{2t}\right)
-\psi\left(\frac{1}{2} \right)\,,
\label{Q}
\end{eqnarray}
where $t=T_c/T_{c0}$ with $T_{c0}$ being the critical temperature of the clean material given in Eq.\,(\ref{Tc}). The scattering parameter 
\begin{eqnarray}
\rho_{0}=  \frac{1}{2\pi T_{c0}\tau}\,. 
\label{rho_0}
\end{eqnarray}
 One can easily rearrange $P$ to the form: 
\begin{eqnarray}
 P =  \frac{1}{\tilde\lambda} -
\ln t -Q\,.
\label{P1}
\end{eqnarray}
Next, one solves the quadratic Eq.\,(\ref{eqP}) for $P$ and chooses the
smaller of two roots (with the minus sign in front of the square root).
Denoting this root as $P_r({\hat\lambda},\rho_0,t)$ where ${\hat\lambda}$
stands for the set of all coupling constants and of partial densities of
states, we obtain an implicit equation for $t(\rho)$ that can be solved numerically:
\begin{eqnarray}
   \frac{1}{\tilde\lambda} -
\ln t -Q(t,\rho)=P_r({\hat\lambda},\rho,t)\,.
\label{t(rho)}
\end{eqnarray}

For the case  $\eta=0$, the only root of Eq.\,(\ref{eqP}) is
\begin{eqnarray}
  P_r=\frac{1-Q(n_1\eta_1+n_2\eta_2)}{n_1 \lambda_{11}   + n_2
\lambda_{22}} \, .
    \label{root}
\end{eqnarray}
Since in this particular case $\tilde\lambda =n_1 \lambda_{11}   + n_2
\lambda_{22} $, we obtain:
\begin{eqnarray}
  -\ln t=Q\left(1-\frac{ n_1^2 \lambda_{11}+2n_1n_2 \lambda_{12}+n_2^2 \lambda_{22}}{n_1 \lambda_{11}   + n_2\lambda_{22}} \right)\,.
\label{supression_old}
\end{eqnarray}
One can verify that this coincides with the suppression formula obtained within the model with factorizable coupling potential,  see Ref.\,\onlinecite{Kogan2002} or a more general work by Openov. \cite{Openov} With the parentheses on the right-hand side equal to 1, this is just the  Abrikosov-Gor'kov result for the $T_c$ suppression by a pair-braker with the scattering parameter $\rho_0$. Thus,  only if $n_1^2 \lambda_{11}+2n_1n_2 \lambda_{12}+n_2^2 \lambda_{22}\le 0$, or   
\begin{eqnarray}
\lambda_{12}\le -\,  \frac{ n_1^2 \lambda_{11} +n_2^2 \lambda_{22}}{ 2n_1n_2}  \,,
\label{supression_old}
\end{eqnarray}
$T_c$ drops to zero at a finite $\tau$. Otherwise $T_c(\rho_0)$  is a decreasing function at all $\rho_0$.   

One can show numerically that  these  features of the  $T_c$ suppression are qualitatively the same for a general two-band case: unlike formulas of preceding sections describing  clean materials, the sign of the interband coupling $\lambda_{12}$ does affect the $T_c$ suppression.  One can verify that the interband scattering causes faster decrease of $T_c$ if the  interband coupling  is repulsive, $\lambda_{12}<0$.

\begin{figure}[t]
\includegraphics[width=8.cm]{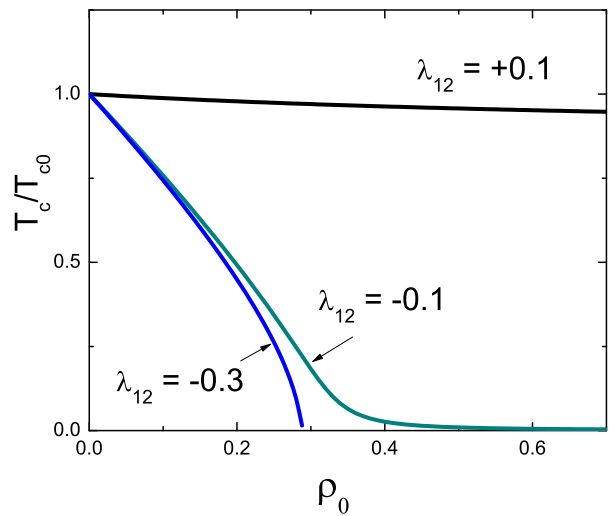}
\caption{(Color online) Suppression of the critical temperature by non-magnetic impurities in a two-band superconductor. The intraband couplings $ \lambda_{11}=0.3$,  $ \lambda_{22}=0.2$ are the same for all curves. The suppression is much stronger when the sign of $ \lambda_{12}$ changes to negative (repulsive). For a strong repulsive  $ \lambda_{12}$, $T_c$ turns zero at a finite scattering parameter $\rho_0$.  
}
\label{fig7}
\end{figure}

To illustrate this point we calculate suppression of $T_c$ 
with the coupling parameters 
$ \lambda_{11}=0.3$,  $ \lambda_{22}=0.2$, and with three different values of $ \lambda_{12} $ shown in  
Figure \ref{fig7}. Whereas with positive
$\lambda_{12}$ the suppression is weak and similar to the case of 
materials having  one anisotropic gap, the suppression for
$\lambda_{12}<0$ is much stronger. This finding can be checked
experimentally and, in fact, the published data on unusually fast
suppression of $T_c$ with carbon, aluminum, or lithium 
doping\cite{Karp} imply that MgB$_2$ might have  repulsive interband coupling. Otherwise, it is
hard to reconcile the $T_c$ suppression by a factor of 4 by 15\% of
C substitution. If one interprets this effect as caused by impurities scattering, Eq.\,(\ref{t(rho)}) with $\lambda_{12}=+0.06$ provides $\rho_0\sim 10^3 $ needed for such a suppression. This value of the scattering parameter corresponds to unrealistically short men-free path $\ell\sim 1\,$\AA or less. Changing the sign of $\lambda_{12} $ to negative (i.e. taking the interband coupling as repulsive), results in $\rho_0\approx 0.37 $ and a reasonable estimate of $\ell\approx 400\,$\AA.  
The negative interband coupling would imply opposite signs of the order parameter on the  two effective bands of MgB$_2$, i.e. $\pm $s  order
parameter, a proposition calling for more studies. 

\section{Summary}

We have presented a two-band weak-coupling $\gamma$-model that  takes into account  self-consistently all relevant coupling constants to evaluate temperature dependencies of the two gaps, of the superfluid density,
and of the specific heat in clean s-wave materials. The interband coupling is shown to have a strong effect an these dependencies irrespective of the sign of this
coupling. In particular, if the interband coupling is negative (repulsive) it may cause the two order parameters to have opposite signs, i.e. the order parameter may have the $\pm$s structure. In this case, the  $T_c$ suppression by interband scattering  should be very strong, the feature that can be utilized as a signature of the $\pm$s order parameter. We speculate that a strong $T_c$ suppression by various dopands in MgB$_2$  may signal such a possibility. 
All these features make the model advantageous to the
empiric and \textit{not self-consistent} $\alpha$-model commonly employed to interpret the data on penetration depth and specific heat of two-gap materials. 

\acknowledgements

We thank R. T. Gordon and H. Kim for help with the experiments, J. Karpinski for MgB$_2$ and D. K. Christen for V$_3$Si single crystals,  P.~C.~Canfield, A.~Carrington, A.~V.~Chubukov, S.~L.~Bud'ko, A.~J.~Legett, I.~I.~Mazin, J.~Schmalian, M.~A.~Tanatar, Z. Tesanovic and F. Catus for interest and discussions. Work at the Ames Laboratory is supported by the Department of Energy-Basic Energy Sciences under Contract No. DE-AC02-07CH11358. R.P. acknowledges support of Alfred P. Sloan Foundation.

\end{document}